# Fundamental cause for superior optoelectronic properties in halide perovskites

Xiaoming Wen, Baohua Jia

School of Science, RMIT University, Melbourne, VIC 3000, Australia

**ABSTRACT**

Halide perovskites have emerged as revolutionary materials for high performance photovoltaics and optoelectronics due to their superior optoelectronic properties. The physical origin for the superior optoelectronic properties of halide perovskites so far is still poorly understood. Here we propose and demonstrate a hypothesis that electron upconversion (detrapping) driven by ionic energy reservoir is the fundamental cause for the superior optoelectronic properties of halide perovskites. We fully consider ionic influence on the electronic dynamics in mixed ionic-electronic conduction system by introducing new concepts of ionic energy reservoir, ion-electron coupling and ion-phonon scattering. We clarified that the ionic beneficial effect originates from the different mechanisms from that of the detrimental effect of mobile ion. Our hypothesis consistently interprets that the electron detrapping directly leads to significantly enhanced fluorescence efficiency, prolonged carrier lifetime, and increased diffusion length, as well as the anomalous phenomena of defect healing and defect tolerance, which are responsible for the excellent device performance of halide perovskites. By adding the ion-electron coupling into the rate equations, we establish the physical correlation between electronic dynamics in the timescale of nanosecond-microsecond and ionic dynamics in the timescales of second to hour. This finding adds the missing puzzle into the holistic physics picture and provides a deep understanding of halide perovskites and ion-electron interaction in mixed ionic-electronic semiconductors. Our results suggest the possibility of maximizing the potential of halide perovskite devices through enhancing ion-electron coupling.

Corresponding: Xiaoming Wen xiaoming.wen@rmit.edu.au, Baohua Jia baohua.jia@rmit.edu.au

Topic of Content

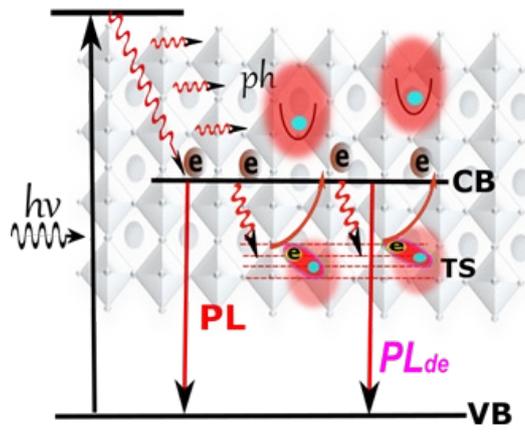

# 1. The dilemma of current concepts in carrier recombination dynamics in halide perovskites

Halide perovskites have emerged as the most promising materials for high efficiency photovoltaics and optoelectronics.[1-3] The exceptional high performances have been widely attributed to their unique optoelectronic properties, such as long carrier lifetime and long diffusion length.[4-5] During the last decade intensive effort has been made for understanding the underlying physics in halide perovskites by various experimental and theoretical techniques. However, the most fundamental question remains elusive: What makes halide perovskites superior to conventional semiconductors? What is the fundamental cause of their superior optoelectronic properties and anomalous phenomena in devices, such as defect healing and defect tolerance? Answering these questions is critical to fully understand the potential of perovskites and design better devices with optimum performance.

One of the anomalous phenomena in halide perovskites is that the photogenerated carriers exhibit exceptionally long lifetime and long diffusion length, much longer than that predicted by the Langevin theory.[6-7] In principle, carrier lifetime consists of radiative and nonradiative components, corresponding to radiative and nonradiative recombinations. Radiative recombination is inherently determined by electron-hole interaction strengthen (wavefunction overlapping). When nonradiative recombinations increase, such as defect trapping, carrier lifetime decreases. Therefore, the maximum observed lifetime appears when the nonradiative recombinations are completely suppressed. In other words, the observed carrier lifetime cannot exceed the inherent radiative lifetime if no other specific physical mechanism involves. In halide perovskites, the observed lifetime is 2-4 orders of magnitude longer than the inherent radiative lifetime.

Many physical microscopic origins were proposed to interpret such a long lifetime and diffusion length, such as delayed fluorescence (through triplet state or trapping-detrapping processes) or indirect bandgap transition,[8-13] formation of polaron and thus screening,[14-16] giant Rashba effect,[16-18] ferroelectric domains,[19-21] photon recycling.[22-24] Basically, these physical origins can partly explain the observed long carrier lifetime to a certain extent but their rationality is experimentally disproved and thus to date are not extensively accepted.[25-26]

Regarding the origin of delayed fluorescence (type-I thermal activated),[27] it cannot present consistency between the electronic structure (singlet and triplet states), composition and observed carrier lifeitm.[28-29] For the origin of polaron formation, it cannot present consistent interpretation for the long carrier lifetime and some open questions are hard to be addressed experimentally, such as effective mass, binding energy, coherence length of polarons, the interaction between A-site cation and inorganic sub-lattice, the ultrashort formation time of polaron in the sub-picosecond timescale and relationship with detailed compositions.[14, 16, 30-31] In terms of the origin of Rashba band splitting, a spin-forbidden nature of interdomain recombination can reasonably explain the long carrier lifetime in $MAPbI_3$, due to spin-forbidden indirect transition thus slow bimolecular recombination.[17, 32] However, nowadays it is difficult to verify the spin-forbidden and spin-allowed recombinations occurring with and without Rashba splitting, respectively, by suitable spectroscopic measurements.[18, 26] Temperature dependent observations of carrier lifetimes contradicts the origin of ferroelectric domain because ferroelectric domains are expected to disappear at high temperatures, which enable faster recombination of photogenerated carriers.[33-34] The slow carrier lifetime is ascribed to the repeatedly reabsorption-reemission in the theory of photon recycling.[35] However, there is an apparent problem when interpreting the superior transport properties of hybrid perovskites over the inorganic ones because photon recycling requires photoluminescence quantum yield (PLQY) of the material to be close to unity but PLQY in $MAPbI_3$ is low.[36-37]

Most importantly, these interpretations are all based on electronic dynamics and mostly ignore the ionic contributions. The most anomalous phenomena in halide perovskite are illumination-induced variation of optoelectronic properties occurring in the timescales of second to hour, which is generally ascribed to ionic relevance. Obviously, these existing theories based on electronic dynamics fail to explain the phenomena of illumination-induced fluorescence enhancement and defect tolerance.[38-40] For the observed illumination-induced fluorescence enhancement (also referred to as photobrightening, defect curing, and defect healing, discussed later), it is difficult to interpret why and how light can result in such a slow variation.

Despite intensive efforts over the past decade to understand the fundamental properties and underlying physics of halide perovskites, a complete and consistent interpretation of their superior optoelectronic properties and anomalous phenomena in devices has not been established. Instead, phenomenological extensions have been used to explain the observed effects, such as defect healing and defect tolerance.[41, 40] The influence of ionic contribution on the photogenerated carrier recombination has been largely ignored due to the missing theoretical framework on ion-electron interaction, even though halide perovskites have been confirmed to exhibit mixed ionic-electronic conduction.[42-43] This lack of fundamental understanding of the holistic role ions have been playing in the entire physics picture has greatly restricted the exploration of the true potential of perovskite materials and find the theoretical limit of their resulted optoelectronic devices.

## 2. Ionic effects in halide perovskites

The fundamental difference from conventional semiconductors is that halide perovskites are mixed ionic-electronic semiconductors in which not only charged carriers (electrons and holes) but also ions intimately impact their optoelectronic properties.[44-45] In recent years, the ionic effects in halide perovskites have been intensively investigated by theoretical simulation and limited experiments.[46-49] Perovskites have been confirmed to exhibit mixed ionic-electronic conduction.[45-47, 50-52] The observed anomalous effects, including photobrightening, defect healing and defect tolerance, as well as I-V hysteresis,[45, 53-54] light soaking and phase segregation,[55-57] are suggested to be relevant to ionic effects.[40, 58-59]

Mobile ions have been intensively investigated as a negative effect in halide perovskites, because it is widely accepted that mobile ions are intimately relevant to fluorescence intermittency (blinking), I-V hysteresis, phase segregation, degradation thus instability.[48, 54, 60-66] It needs to be pointed out that there exist positive and negative ionic relevant effects in halide perovskites. The opposite phenomena have been observed in our experiments, ionic induced PL quenching (quenching in ensemble and blinking in single grain) and ionic induced PL enhancement. It is necessary to emphasize that the positive effect and negative effect are the different physical effects, rather than the different aspects of the same physical effect.

To date, the beneficial ionic effects, such as illumination induced PL enhancement (photobrightening), defect healing/curing and defect tolerance, memory effect, have been much less understood compared to ionic detrimental effects and electronic effects in halide perovskites, mostly due to the complexity of ions in halide perovskites. These effects generally exhibit extremely slow dynamics in timescale of seconds and are suggested to relevant to ionic-electron interaction. There is no established physical model to describe the dynamic behaviours of ion-electron interaction so far.

On one hand, it lacks effective technique for selectively probing ions from overlapped ionic-electronic effects. The ions do not have characteristic absorption and emission like electronic transitions, thus conventional spectroscopic techniques are not directly applicable. On the other hand, ions can be multiple species that are correlated with different elemental atoms either positively or negatively charged, mobile or immobile. Ions in halide perovskites include various species, which are difficult to clearly determine. In the meantime, ions exhibit multiple roles, thus leading to different physical effects in halide perovskites. These roles can be basically classified into four aspects: (1) charge effect: it has been reported that the mobile ions in halide perovskites most likely include interstitials such as $I^-$, $Br^-$, $MA^+$, $Pb^+$, $FA^+$ and vacancies such as $V_I$, $V_{MA}$, $V_{Pb}$.[44, 67-68] For charges in the lattice, Coulomb effect induced lattice distortion is expected, and screening effect of electric field will occur upon ion accumulation.[69-70] (2) defect effect: as kinds of defects, such as interstitial and vacancy, mobile ions can originate from either the initial fabrication or activation by illumination or electric field. These defects may act as recombination centres thus impact electron-hole recombinations, particularly the Shockley-Read Hall recombination (defect trapping), through enhanced ion-phonon and electron-phonon coupling.[5, 61, 67-68, 71-73] (3) slow response effect: ions have much larger mass than electrons. Their diffusion is essentially restricted by both the potential of the lattice and occupancy of corresponding lattice. Up to 10 μm diffusion - halide substitution has been confirmed in mixed halide perovskite film under illumiation.[46] Ions exhibit a much lower mobility, and a much slower time response generally in tens of ms or longer.[44, 59, 62, 74-76] (4) Composition effect: mobile ions can be generated when ion escape from the lattice and diffuse under illumination or electric field. They can further substitute with the atoms in the lattice or accumulate spatially at the interface, which generally results in decomposition, degradation, or phase segregation.[55, 57, 77-80] Moreover, these ionic effects can directly or indirectly cause instability through degradation, decomposition, phase segregation, or induce lattice strain, energy band alignment, defect formation and electric screening.[47, 58, 77, 81] These simultaneous effects as well as the multiple species and roles of ions significantly increase the complexity of ion-electron interaction. To date, the theory describing ion-electron interaction is not yet established,[46-47] the direct experimental observation has been hindered and therefore, the critical role played by ion-electron interaction in perovskite has not been properly revealed and recognised.

## 3. Immobile ion and mobile ion

Halide perovskites has a crystal structure following the formula $ABX_3$ and there are three ions in the lattice of perovskites: A-cation, B-cation or halide anion. Under external excitation of illumination or electric field, two species of ions can be generated: the ion localized in the lattice as immobile ion and ion out of lattice as mobile ion. Under illumination the ion can obtain extra energy through ion-phonon scattering and exhibit higher ionic temperature, referred to as hot ion. The hot ion can be localized in the lattice (immobile), dynamic and nonequilibrium, acting as ionic energy reservoir, taking into account the long lifetime, which can provide extra energy to the lattice. In this case ion is localized in the lattice so that the unit cell displays as charge neutrality. It is expected that the effects originating from charge and defect are suppressed for such 'immobile ion'; the effects from its lattice (slow response, strains, and composition) can be displayed. The inefficient thermal transport is ascribed to the suppressed acoustic phonon relaxation in halide perovskites.[82-84]

The second specie ion forms when ion (A-cation, B-cation or halide anion) escapes the lattice binding during illumination. With illumination the ion obtains extra energy (increased potential) so that it has increased possibility to escape the lattice binding, generating interstitial and vacancy.[67, 72, 79, 81] These mobile ions are charged and can act as defects or trapping/recombination centre.[48, 85] In this case the effects from charge and defect are enhanced, for example, PL quenching or blinking.[62, 86]

It is necessary to note that two ionic effects occur simultaneously under illumination because generation of mobile ion is randomly in microscopic lattice and the possibility is determined by both the potential of the ion and ionic binding energy (formation energy). Under illumination at low intensity ion obtains less extra energy with lower potential due to low density of hot phonons, thus lower possibility to generation of mobile ion. The dominant effect will originate from localized hot ion. Therefore, PL enhancement with prolonged lifetime can be observed. In contrast, at high intensity of illumination, ion has the higher possibility to escape the lattice binding as mobile ion. The effects from charge and defect will be dominant. Consequently, PL blinking or quenching is usually observed at high intensity of illumination.

The term "ion" is conventionally understood as mobile ion (moving charge), which only considers their charge effects and ignores other important effects. It does not reflect the overall ionic effects in halide perovskites. In this manuscript, we extend the ionic concept to a generalized ion with holistic effects, including charge, large mass and size than an electron, defects, long lifetime typically in seconds and low mobility. Therefore, here the ion includes multiple physical meanings and effects, rather than just the conventional moving charge. With such generalized ionic concept, the observed ion-correlated effects with much slower response times (long lifetime) lead us to speculate that energetic ions can store energy, which we define as an "ionic energy reservoir".

Under such generalized concept and in-depth analysis of the volumetric data on the observed but yet explainable unique features of halide perovskites, we propose a hypothesis that electron detrapping (upconversion) by ion-electron interaction is the fundamental cause for the superior optoelectronic properties and excellent device performance in halide perovskites over conventional semiconductors. Considering halide perovskites have become a large family, here we focus on representative $MAPbI_3$, $FAPbI_3$ and their mixture, which are intensively investigated and applied in various devices. The findings are applicable to the other halide perovskites in general. We will demonstrate that such electron detrapping directly results in significantly increased fluorescence efficiency, prolonged carrier lifetime and diffusion length, as well as defect healing and defect tolerance. As a consequence, electron detrapping significantly decreases energy loss in the system and results in dramatically improved performance of perovskite devices. Electron detrapping can also consistently interpret the observed series anomalous phenomena in halide perovskites, in particular ionic correlated slow effects. This hypothesis may open the door to understand the long-missing fundamental ion-electron interaction, which is crucial for both new scientific discoveries and technological breakthroughs.

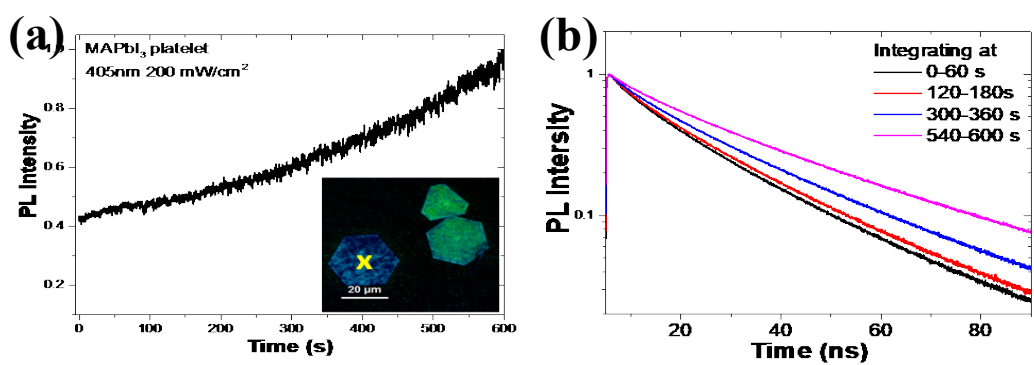

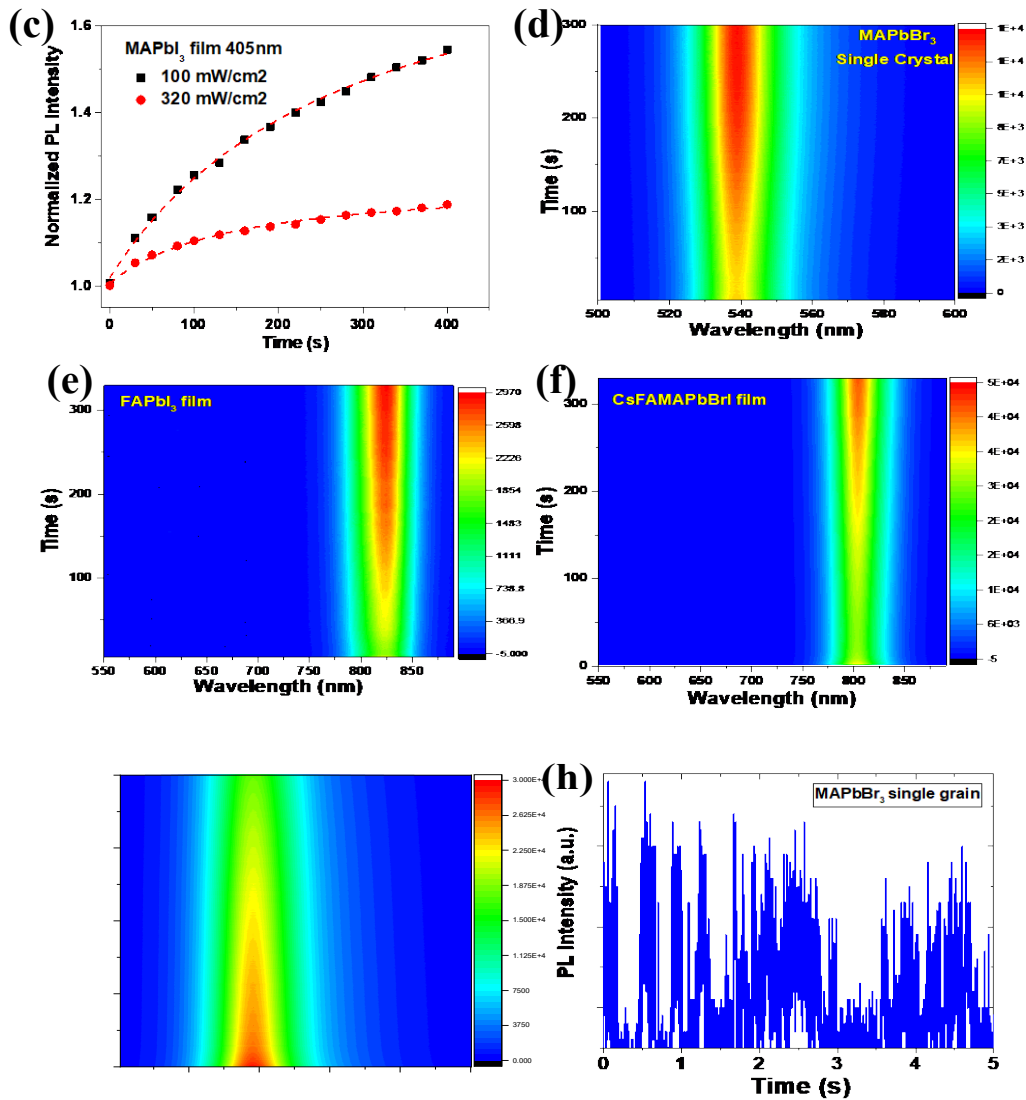

Fig. 1 illumination induced fluorescence enhancement and increased carrier lifetime in timescale of seconds in halide perovskites. PL Intensity (a) and PL decay curves (b) in MAPbI$_3$ nanoplatelet under continuous illumination (405 nm 200 mW/cm$^2$). Inset (a) is the PL imaging of the sample. (c) PL intensity evolution under continuous illumination in MAPbI3 film by spin coating. PL spectra as a function of illumination time in (d) a single crystal MAPbBr$_3$ (e) FAPbI$_3$ film (f) mixed CsFAMAPbBrI film under constant illumination intensity (405 nm at intensity of 120 mW/cm$^2$). (g) PL spectrum is observed in MAPbBr$_3$ single crystal under 405 nm excitation at a fluence of 350 mW/cm$^2$. (h) Fluorescence intermittency in MAPbBr$_3$ single grain as a function of time under the continuous excitation of 405 nm.

## 4. Electron detrapping (upconversion) in halide perovskites

Illumination-induced fluorescence enhancement and quenching, also refer to as light soaking, have been widely observed in halide perovskites.[38-39, 44, 59, 61-62, 87-88] The phenomenon of illumination induced fluorescence enhancement is also described or primarily interpreted with different terms, such as photobrightening, defect curing, defect healing, illumination induced PL enhancement. In the meantime, the opposite effect, illumination-induced fluorescence quenching and fluorescence intermittency (blinking) are also observed in halide perovskites,[62, 89-91] exclusive degradation and decomposition. To date, all proposed interpretations are only phenomenon descriptions and/or primary explanations; the underlying physical mechanism is still unclear.

Illumination induced fluorescence enhancement can be typically expressed as increased intensity of either PL integral intensity or PL spectra, as a function of illumination time. Fig. 1a shows examples of illumination-induced fluorescence enhancement as a function of illumination time in $MAPbI_3$ (MA= $CH_3NH_3$) perovskite nanoplatelet fabricated by chemical vapor deposition (CVD). The details of the nanoplatelets are reported in reference[92] and also shown in Supporting Information (SI). With continuous illumination (405 nm laser of 200 mW/cm$^2$), PL enhancement is clearly observed (Fig. 1a) expressed by the integral intensity. The carrier lifetime accordingly exhibits significant increase (Fig. 1b), consistent with the previous observations in $MAPbI_3$ film by spin coating.[38, 87] The same phenomenon can be observed by continuously acquired PL spectra in $MAPbI_3$ film by spin coating (1c), $MAPbBr_3$ single crystal (1d), $FAPbI_3$ polycrystalline film (1e) and $(FAMA)PbI_3$ polycrystalline film (1f) under continuous illumination (constant excitation). The similar phenomenon, also termed as photobrightening, has been widely observed in various halide perovskites with different fabrication methods, such as inorganic-organic hybrid halide perovskites, inorganic perovskites, halide mixed perovskites, $MAPbI_3$, $MAPbBr_3$, $FAPbI_3$, $CsPbBr_3$.[82, 93-94] The apparent carrier lifetime is found to simultaneously increases (Fig. 1b), which has been usually ascribed to defect healing or defect curing.[38, 41, 87] It is a prerequisite for these experiments to exclude illumination-induced degradation, decomposition, and phase segregation, which can be confirmed by the recovery of photoluminescence intensity and the absence of any detected $PbI_2$ through sensitive spectroscopic analysis.[95-98] The problem with the aforementioned models is that the proposed electronic transition cannot account for such reversible variations induced by illumination.

In addition to ion-induced PL enhancement, it is necessary to note there exists the opposite effect in halide perovskites, mobile ion induced PL quenching.[38, 61, 99-101] Fig. 1g shows the PL spectrum of $MAPbBr_3$ single crystal (the same sample as that in Fig. 1d) under the same wavelength excitation but with a higher fluence of 350 mW/cm$^2$. In other words, at low excitation intensity it exhibits PL enhancement, in contrast, at higher excitation intensity it exhibits PL quenching (intensity decrease). Fig. 1h shows the PL intensity in a single grain of $MAPbBr_3$ film as a function of time under the continuous excitation at 405 nm, which presents as fluorescence intermittency (also refer to as blinking).[44, 62, 102-105]

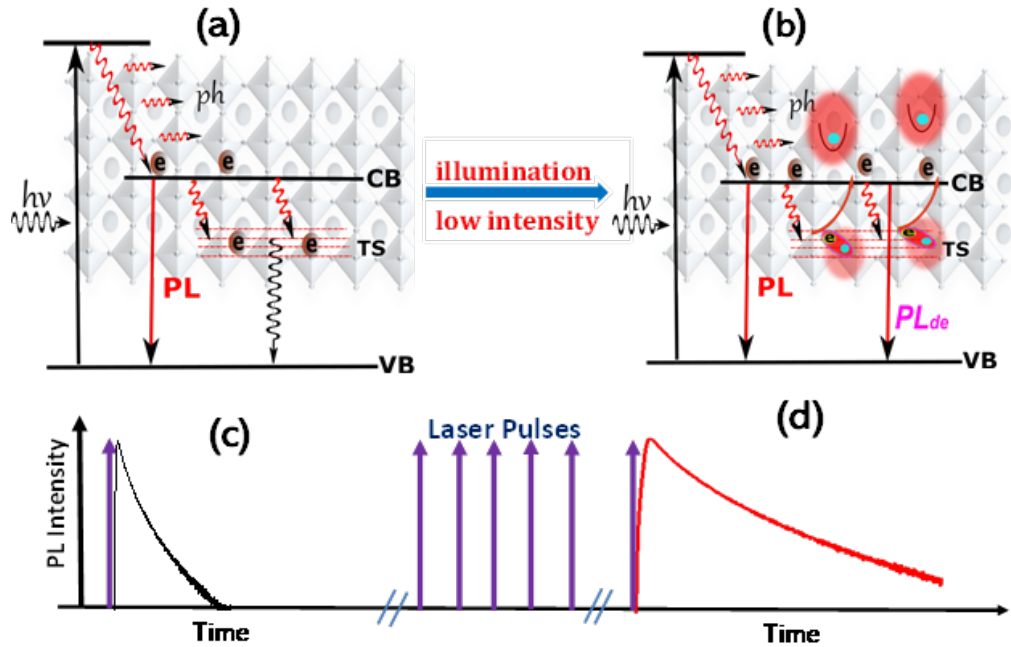

Fig. 2. Schematic electron detrapping by ion-electron coupling. Ph-phonon, CB-conduction band, VB-valance band, TS-trapping state. (a) without ionic activation. (b) upon illumination the ions are activated as energetic ions. By ion-electron coupling the trapped electrons can be detrapped back to the conduction band. The corresponding PL decay curves for non-detrapping (c) and with detrapping (d), respectively.

To understand the underlying physics of this anomalous effect, we propose a hypothesis of electron detrapping through ion-electron coupling, as schematically elucidated in Fig. 2. It should emphasize that essential difference between halide perovskites and conventional semiconductors is the feature of mixed ionic-electronic conduction in halide perovskites. Initially ions in the lattice are non-activated and ion-electron coupling is weak and negligible. Thus, the carrier dynamics is determined by the inherent radiative recombination and nonradiative recombinations as competition components, which is the same as that in conventional semiconductors. Upon continuous illumination the photogenerated hot carriers quickly relax to the conduction band dominantly by emitting hot phonons. The hot phonons can dissipate energy as lattice heat in conventional semiconductors. The typical cooling time for the hot carriers is only few picoseconds in solid-state semiconductors and even shorter in organic materials.[106-109] It is intriguing to note that halide perovskites exhibit significant phonon bottleneck due to weak electron-phonon coupling, and the slowed relaxation of acoustic phonons results in acoustic-optical phonon upconversion.[31, 82, 94, 110-112] Under illumination, ions in the lattice are expected to be activated as energetic ions by ion-phonon scattering.[74, 113] Then the carriers in the conduction band relax back to the valence band by radiative and nonradiative recombinations. In the meantime, some photogenerated electrons are trapped by the defects, generally as energy loss in solar cells and optoelectronic devices. Note that ionic effects in halide perovskites exhibit extremely long lifetime, usually in the timescales of second to hours, usually referred to as memory, structural polarization and PL quencher.[38, 46, 75-76, 114-115] With the generalized concept of ions, we propose the energetic ion (hot ion) can store the emitted phonon energy (depleted into the lattice) acting as ionic energy reservoir through enhanced ion-electron coupling due to the ultralow thermal conductivity in halide perovskites.[82-84, 94, 116] The trapped electrons can be detrapped back to the conduction band by obtaining extra energy from the energetic ions (enhanced ion-electron coupling). The processes can be schematically described as:

$$Ion + phonons \rightarrow Ion^* \qquad (1)$$

$$Ion^* + e \rightarrow Ion + e^* \qquad (2)$$

As a consequence, the carrier population in the conduction band is increased, which in turn changes the recombination dynamics, as discussed in detail later. From the energy transfer point of view, the lost energy through phonon emission and trapped electrons is largely regained. It is expected that this effect can significantly improve carrier optoelectronic properties, such as increased apparent carrier lifetime and carrier diffusion length, increased PL quantum yield, as well as defect healing and defect tolerance. These superior optoelectronic properties are well known to be responsible for the high performance of perovskite solar cells and optoelectronic devices.

To obtain the deep insight into the influence of carrier detrapping on the apparent carrier lifetime, it is necessary to recall the measurement of carrier lifetime by time-resolved fluorescence techniques. In the measurement of time correlated single photon counting (TCSPC), the lag time of each emitted photon is obtained by the correlation relative to the excitation pulsed laser. The PL decay curve is obtained by statistically calculating the photon number as a function of the lag time. A smooth PL decay curve can be acquired if the photon number is sufficiently large, although each individual photon emits at a random lag time. Upon a pulsed laser excitation, the carriers are photogenerated in an excited state that is determined by the photon energy. Following an ultrafast cooling, the carriers relax into the conduction band dominantly by efficient phonon scattering. Then the carriers deplete their energy through recombination processes, including defect trapping (phonon assisted recombination, also known as the Shockley-Read-Hall (SRH) recombination), bimolecular recombination and Auger recombination, back to the valence band. Note that a PL decay curve is acquired by integrating a large number of photons thus many cycles of excitation-depletion, generally including $10^6$ to $10^8$ cycles. For other time-resolved spectroscopic techniques, for example ultrafast up-conversion fluorescence and ultrafast transient absorption,[82] a high quality decay curve of fluorescence or absorption can be acquired by integrating a lot of pulses.

One usually considers the energy of photoexcited carriers deplete completely before the next pulse arriving. A PL decay curve represents a completed process of excitation-depletion cycle by assuming all energy of carriers are excited and depleted in one cycle. In the case of halide perovskites, such assumption may not be valid due to carrier detrapping through ion-electron coupling and extremely long ionic lifetime, usually in timescale of seconds, that is 6-8 orders of magnitude slower than that of electrons, which is in the ns timescale.[58-59, 117] Therefore, during the measurement, ions can accumulate energy essentially by carrier-phonon coupling. Some ions can be activated to become energetic ions or to escape the lattice as mobile ions and defects for recombination centres, such as interstitial and vacancy. These energetic ions can feedback their energy to the trapped carriers by ion-electron coupling, resulting in carrier detrapping.[118-119] In conventional semiconductors the trapped carriers will lose their energy to the lattice and relax quickly back to the valence band, as an energy loss. In contrast, these carriers in halide perovskites can regain energy and jump back to the conduction band by carrier detrapping; eventually contributing to the efficiency of solar cells or optoelectronic devices.

To provide quantitative understanding, we describe the carrier density $n = n(t)$ at the conduction band by a rate equation, taking into account the balanced electron-hole density and free carrier nature in halide perovskites:[120-121]

$$\frac{dn}{dt} = \sigma I_{ex}\delta(t) - An - Bn^2 - Cn^3 \qquad (3)$$

where the first term $\sigma I_{ex}\delta(t)$ corresponds to photogenerated carriers that is relevant to the absorption cross section σ and excitation intensity $I_{ex}$, considering this process is ultrafast as a delta function $\delta(t)$. The terms of $An$, $Bn^2$ and $Cn^3$ signify SRH recombination via sub-bandgap trap states, bimolecular recombination and Auger recombination, respectively.[120]

It is necessary to note that this traditional equation completely ignores the ionic influence, which miss the critical component in the physical picture for halide perovskites. Here we add an additional term to include ionic contribution: carrier detrapping through ion-electron coupling. The overall rate of carrier detrapping will be proportional to the population of energetic ions $n_{ion}$ and carrier density in the trapping states $n_d$ because the detrapping originates from the ion-electron coupling. Assuming the shallow defect state can effectively populate with the conduction band then carrier detrapping from the deep defect state can be accounted as an additional term in the rate equation:

$$\frac{dn}{dt} = \sigma I_{ex}\delta(t) - An - Bn^2 - Cn^3 + K\sigma' n_{ion} n_d \qquad (4)$$

$$\frac{dn_d}{dt} = An - K\sigma' n_{ion} n_d - K_1 n_d \qquad (5)$$

where $\sigma'$ is the cross section of detrapping and $n_{ion}$ is the density of the energetic ions, both are composition and fabrication dependent, $K$ is a constant. $n_d$ is the carrier density of the deep defect state and $K_1$ represents the relaxation rate from the deep defect state into the valence band. As discussed in detailed in SI, halide perovskites have extremely low thermal conductivity, which suggest enhanced carrier detrapping due to higher population of the acoustic phonon.[82, 122-123]

For halide perovskites, it is expected $K_1$ is very small due to the weak electron-phonon coupling, which is further supported by the extremely low thermal conductivity.[98-100] Note the population of energetic ions $n_{ion}$ is dependent on the integral illumination, relevant to intensity, duration and probably wavelength. It has long lifetime typically in seconds, orders of magnitude longer than that of electrons typically in nanoseconds. Therefore, $D = K\sigma' n_{ion}$ is a slowly varying function during carrier cycle of photogeneration-depletion and thus it can be approximately treated as a constant in electronic dynamics. We emphasize that Equation 4 establishes the correlation between electronic dynamics in nanoseconds and ionic influence in seconds, that is, during each electronic cycle from excitation to depletion this equation represents the carrier recombination with the ionic contribution that sensitively depends on the activation of energetic ions ($n_{ion}$). As we show below in the simulation, this additional term can significantly prolong the carrier lifetime; the extent of the prolonging depends on $n_{ion}$ that is determined by the accumulation of illumination (Fig.2 and Fig. S5).

On the other hand, it is not easy to clearly quantify the correlation of detrapping on the prolonged carrier lifetime because Equations 4 and 5 include multiple parameters and detailed dynamic curves vary with the details of parameters as well as initial conditions. According to the simulation in Fig. S5, upon enhanced detrapping, the decay of the deep defect state tends to be similar to that in the conduction band. So far, the detailed parameters for carrier detrapping are unknown (more details in SI). We can consider approximately an extreme condition as the minimum influence on the carrier dynamics: the detrapping is nearly as effective as trapping $n_d \propto n$ and the relaxation of defect carriers is negligible to the valence band, so that the detrapping term can be approximately expressed as $K\sigma' n_{ion} n$. We can simulate the carrier dynamics as the influence of carrier lifetime by the detrapping by:

$$\frac{dn}{dt} = \sigma I_{ex}\delta(t) - An - Bn^2 - Cn^3 + K\sigma' n_{ion} n \qquad (6)$$

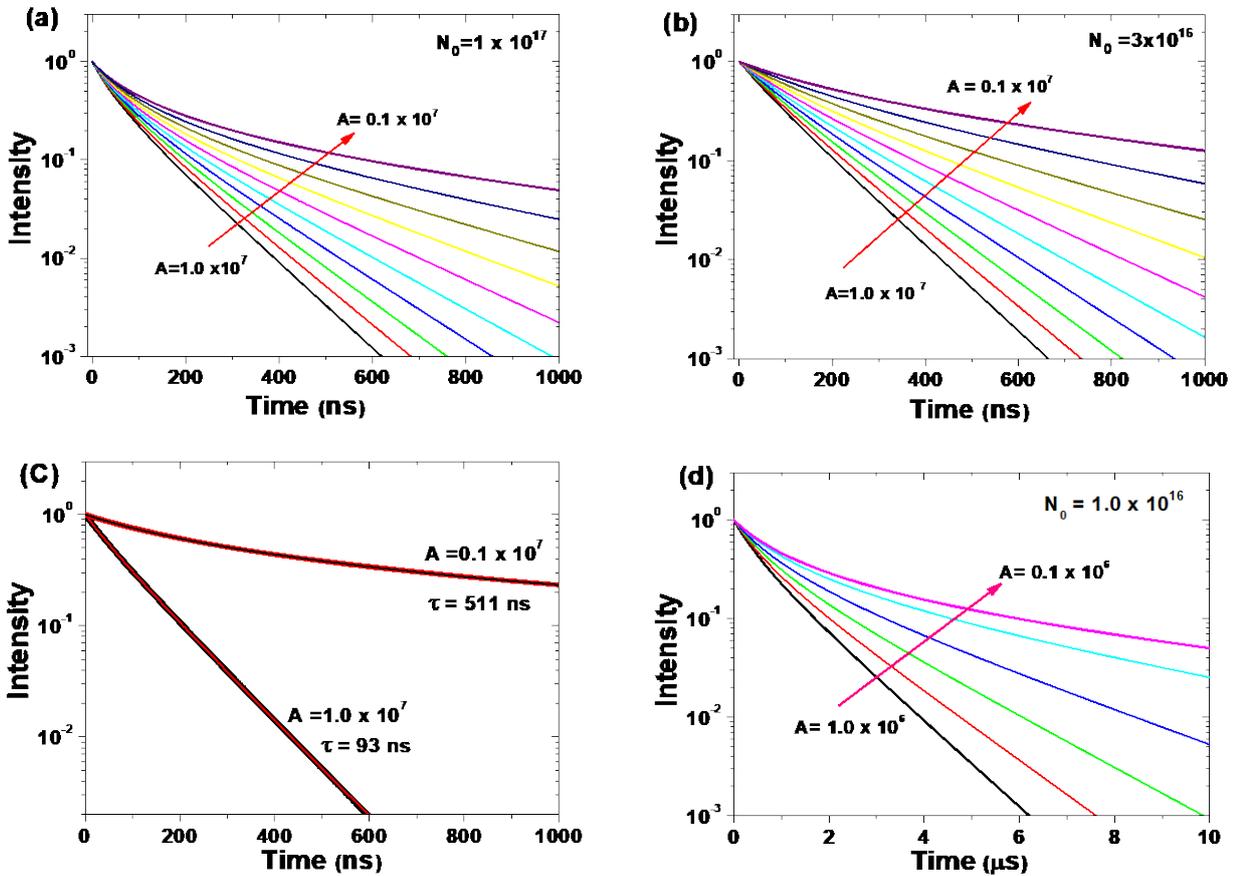

Fig. 3. Simulation of the PL decay curves by rate equation (6): with increasing extent of detrapping (A from 1.0 ×$10^7$ decreases to 0.1×$10^6$), the apparent carrier lifetime significantly increases. (a) at carrier density of $10^{17}$ cm$^{-3}$ and (b) at carrier density of 3×$10^{16}$ cm$^{-3}$. Simulation using typical parameters of B=$10^{-10}$ and c=$10^{-28}$. (c) The average lifetimes increase from 93 ns at $A' = 1.0 \times 10^7$ to 511 ns at $0.1 \times 10^7$. (d) for the case of single crystal $A' = 1.0 \times 10^6$ to $0.1 \times 10^6$ at carrier density of $10^{16}$ cm$^{-3}$, carrier lifetime significantly increases from 822 to 2500 ns.

Under continuous illumination, the density of energetic ions increases. Therefore, ion-electron coupling induced detrapping can be enhanced, which results in the deceased term of $A' = (A - K\sigma'n_{ion})$. In the simulation by Equation 6, with decreasing $A'$ the carrier lifetime significantly increases (Fig. 3a and 3b). The average lifetimes are obtained by two-exponential function fitting and calculated by $\tau = (A_1\tau_1^2 + A_2\tau_2^2)/(A_1\tau_1 + A_2\tau_2)$. As showed in Fig. 3c, the carrier lifetimes increase from 93 ns to 511 ns at a carrier density of 3×$10^{16}$ cm$^{-3}$. Obviously, the macroscopic effect of carrier detrapping is the suppressed defect trapping, which leads to a much longer apparent carrier lifetime than that was predicted by the Langevin theory.[6-7] Essentially, carrier lifetime is determined by the competition between radiative and non-radiative recombinations $1/\tau = 1/\tau_{rad} + 1/\tau_{non}$. The radiative component is intrinsically determined by the wavefunction overlapping of electron and hole, which represents the maximum lifetime with completely suppressed nonradiative recombination. Therefore, it is difficult to rationalize the observed long carrier lifetime if ignoring the ionic contribution, taking into account the defect trapping in perovskites.

Basically, A is the SRH term that is proportional to the density of defect trapping states. When A is small, we find a slower carrier lifetime, corresponding to a lower defect density. In the case of single crystal perovskites with a much lower defect density, extremely longer lifetime is expected. In a simulation for $A' = 1.0 \times 10^6$ to $0.1 \times 10^6$ at a carrier density of $10^{16}$ cm$^{-3}$, the carrier lifetime significantly increases from 822 to 2500 ns (Fig. 3d). It should be noted that this simulation is well consistent with the observations of prolonged carrier lifetime with continuous illumination.[38, 87]

The carrier diffusion length is calculated by $L = \sqrt{D.\tau}$, with $D$ as the diffusion coefficient and $\tau$ as the carrier lifetime.[124-125] Carrier diffusion coefficient of halide perovskites is basically determined by the mobility $\mu$ and temperature $T$: $D = \mu\left(\frac{K_BT}{e}\right)$, where $k_B$ and $e$ are the Boltzmann constant and electric charge of an electron, respectively. The mobility of halide perovskites depends basically on the composition, lattice, impurity concentrations, defect concentration, temperature, and electron and hole concentrations.[126-127] Ion-electron coupling can slightly change the carrier population in the conduction band and significantly change carrier recombination and the carrier lifetime. In this regards, ion-electron coupling will not change *p*- or *n*-type nature but only carrier population in the conduction band. When discussing carrier diffusion, the electron-ion coupling has minor impact on the diffusion coefficient but significantly impact on the diffusion length through prolonged carrier lifetime. Following the above simulation, we can calculate the corresponding increase of carrier diffusion lengths. For solution fabricated MAPbI$_3$ thin film, using electron diffusion coefficient *De*=0.017

cm$^2$/s,[125] the corresponding diffusion lengths are obtained by $L_e = \sqrt{D_e \cdot \tau}$, increasing from 398 nm to 932 nm, corresponding to lifetime from 93 to 511 ns. For a MAPbI$_3$ single crystal *De*=2.05 cm$^2$/s [128] the diffusion lengths increase from 12.98 to 22.64 μm, corresponding to lifetime change from 822 ns to 2500 ns.

The energetic ions generally exhibit a very long response time, also termed as the memory effect,[129] which is usually considered to correlate with the I-V hysteresis and light soaking in perovskite solar cells. To understand the correlation between the memory effect and the slow ion lifetime, we consecutively measured the PL decay curves by TCSPC for an integration time of 30 s under identical condition, except using a continuous wave (CW) illumination as bias, as schematically shown in Fig. 4a. The samples used is MAPbI$_3$ thin film fabricated by spin coating, the detailed sample fabrication and TCSPC measurement are presented in SI.[38] In this study, ps pulsed laser is used as excitation at a wavelength of 405 nm and a repetition rate of 1 MHz. The CW illumination is a xenon lamp through a bandpass filter of 460±20 nm. As shown in Fig. 4b, the three consecutive PL decay curves are (1) excited by the 405 nm laser, (2) 405 nm laser plus CW bias, and (3) removing the bias.

In this experiment, the CW bias generates a constant density of carriers $n_0$. The carrier density is $n'(t) = n_0 + n(t)$. The PL decay significantly accelerates upon adding the CW bias. This is ascribed to the increased weight of bimolecular recombination due to increased carrier density. When removing the bias, a significantly slower PL decay curve is observed than the PL decay initially excited only by the 405 nm laser (Fig. 4b), although the excitation and detection are identical. It should be noted, an identical PL decay curve should be observed if only the electronic transition is considered. This experiment clearly confirmed illumination can result in significantly increased carrier lifetime and such variation exhibits very long lifetime, typically in seconds to hours. The significantly increased PL lifetime can be ascribed to electron detrapping due to well activated energetic ions and thus enhanced ion-electron coupling by the continuous illumination. The extent of ion activation is closely related to the illumination intensity.

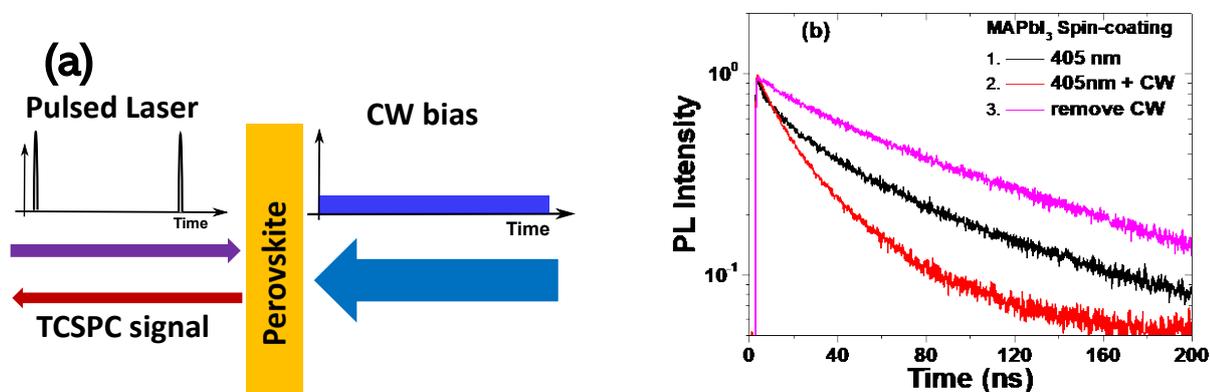

Fig. 4 (a) The configuration of TCSPC measurement combining pulsed laser as excitation and CW illumination as bias. (b) PL decay curves with consecutive TCSPC measurement with (1) pulsed laser only, (2) plus additional CW bias and (3) remove the CW bias in a MAPbI$_3$ film.

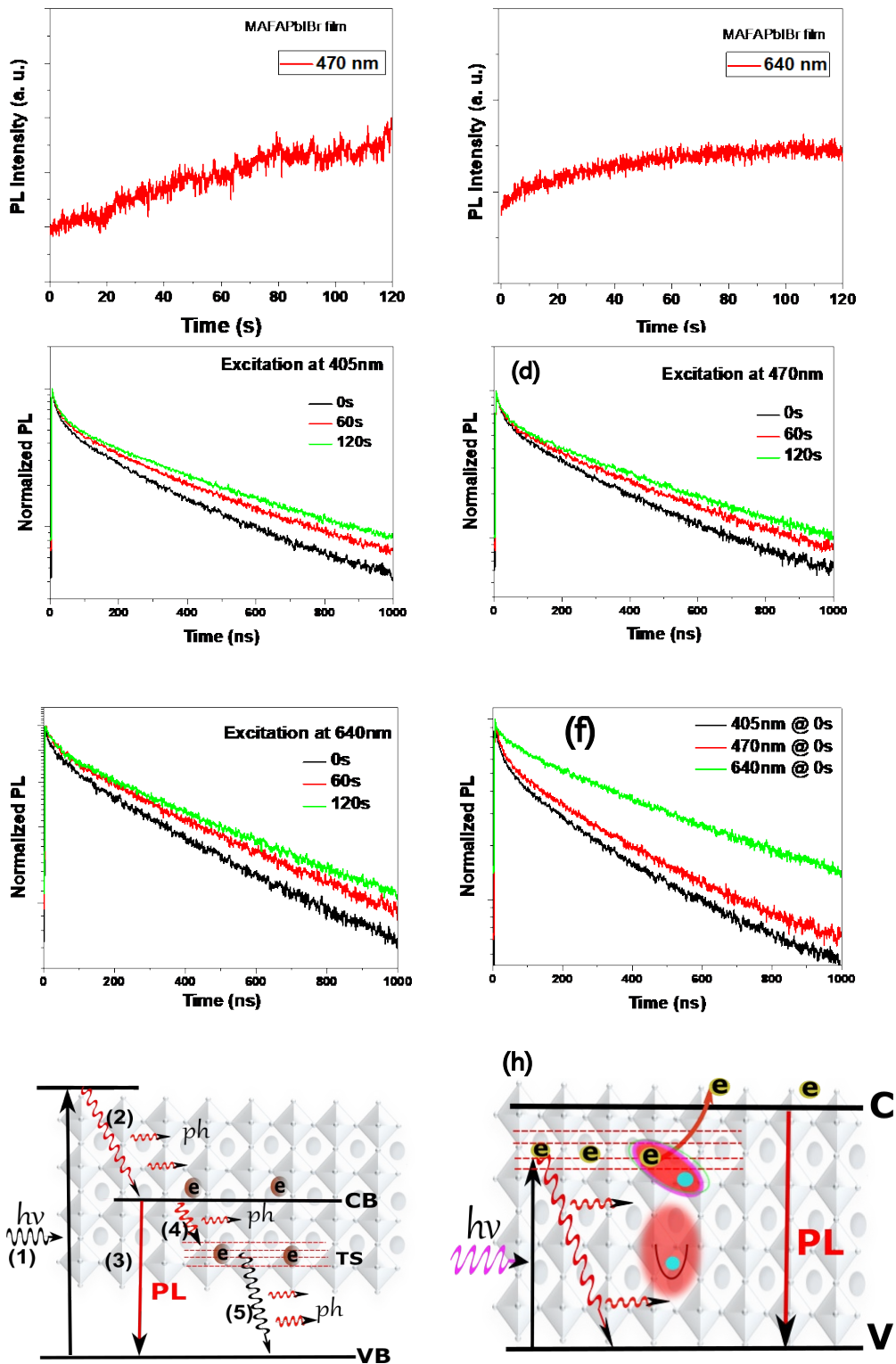

Fig 5. Illumination induced PL enhancement using (a) 470 nm and (b) 640 nm. PL decay curves as a function of excitation time under (c) 405 nm, (d) 470 nm, (e) 640 nm at 0, 60 and 120 s. (f) the comparison of PL decay curves between 405, 470 and 640 nm excitation. (g) General carrier relaxation processes upon excitation for above-bandgap excitation. (h) Conceptual interpretation for the anomalous upconversion fluorescence in halide perovskites.

# 5. Illumination induced PL enhancement at different wavelengths

The illumination induced PL enhancement is performed using different wavelengths. As shown in Fig 5a and 5b for illumination-time-dependent PL experiment on a (FAPbI$_3$)$_{0.85}$(MAPbBr$_3$)$_{0.15}$ [99], with 470 and 640 nm illumination the PL intensity exhibits similar enhancement with that of 405 nm. Moreover, their PL decay curves are measured using TCSPC technique as a function of illumination time. Fig 5c, 5d and 5e show the PL decay curves at 0, 60 and 120 s for 405, 470 and 640 nm excitation, respectively. with continuous illumination the PL decay curve at each wavelength excitation exhibits decreased lifetime, consistent with the PL enhancement. Fig 5f compares the PL decay curves with the excitations of 405, 470 and 640 nm. The different lifetimes are ascribed to the different absorption coefficient and thus absorption depth consider the surface influence.[92, 99] This experiment indicates that not only hot carrier cooling, but also other carrier relaxation processes can contribute phonon for ion-phonon coupling.

Upon excitation the photogenerated electron is excited to the excited state depending on the photon energy[44, 130], Fig 5g. Then the excited electron releases the energy by various processes, such as hot carrier cooling to the conduction band (2), phonon assisted recombination to the trapping states (4, SRH recombination), relaxation from the trapping states back to the valence band (5) as well as band-to-band radiative recombination (3, photon emission), here ignoring Auger recombination considering low density excitation[82, 93]. In addition to the phonon emission (hot phonon LO and transfer to LA to the lattice) in the process of 2, the processes of 4 and 5 can also emit LA phonon coupling with lattice.[82, 93, 113, 120, 131] We note that phonon assisted electron relaxation can result in various carrier dynamic processes, for example, hot carrier cooling from the higher level to the conduction band, defect trapping from the conduction band to the trapping states, carrier depletion/relaxation from the trapping states back to the valence band. Therefore, ion in the lattice can obtain energy during the processes of 4 and 5. For UV and blue excitation, hot carrier cooling can be the major source for LA phonon emission (the thermal energy) coupling into ionic energy reservoir. However, for longer wavelength excitation, for example 470 and 640 nm, ionic energy reservoir can obtain energy from the other LA phonon scattering.

Recently, in all-inorganic lead halide perovskite nanocrystals CsPbA$_3$ (A=I, Br, Cl or mixture of I-Br or Br-Cl), Xiong *et al* observed anomalous upconversion fluorescence, that is, band to band fluorescence is generated by the excitation of below bandgap photon[132], with an exceptionally large energy gain of up to ~ 8 $k_BT$ ($k_B$ is Boltzmann constant, $T$ is temperature).[133] In their experiment the conventional mechanisms of upconversion PL, two-photon excitation and sequential absorption of multiple-photon, can be safely excluded.[57, 131] This observation strongly suggested that the extra energy for the upconversion PL originates from ion-electron coupling in mixed ionic-electronic conduction system.

The upconversion fluorescence can be interpreted by the proposed ion-electron coupling, as shown in Fig. 5h. The incident photons with energy below the bandgap can be partly absorbed[134-135] and the energy will deplete through phonon emission. The active phonons can effectively couple with ions in the lattice to generate energetic ions as ionic energy reservoir. Through enhanced ion-electron coupling, the electrons in the defect states (metastable states) can be upconverted to the conduction band, which results in upconversion fluorescence as band-to-band transition. So far, the detailed ion-phonon interaction is not yet understood with rare reports and the theory has yet established.[113, 136] The similar upconversion fluorescence has also been confirmed in our experiment in neat perovskite films. Energy transfer between ion and electron was theoretically and experimentally investigated in semiconductors, but very little knowledge has been known in halide perovskites so far.[137-139] It should be noted that enhanced ion-electron coupling occurs in localized ion with charge neutrality and mobile ion produce the opposite effect as detrimental effect, which means the method to suppress ionic migration does not contradict with ion-electron coupling. Further comprehensive physical understanding for ion-electron coupling nevertheless is urgently demanded.

Temperature dependent fluorescence has been widely used for phonon relevant study because phonon coupling can be mitigated or suppressed at low temperatures.[27, 34, 140] We performed time-dependent PL spectrum under constant illumination at 200 K in order to safely avoid phase transition, which occurs around 170 K.[34] At low temperature the PL intensity efficiency increases due to suppresses phonon-assisted defect trapping and the PL bandwidth decreases due to mitigated phonon coupling.[141-142] A similar PL enhancement is clearly observed with continuous illumination.

With decreasing temperature phonon couplings are expected to be mitigated or suppressed, which will result in complicated multiple effects in halide perovskites, such as electron-LO phonon scattering, electron-LA phonon scattering, LO-LA conversion, and acoustic phonon-ion scattering. These effects will impact the observation of time-dependent PL spectra: (1) Phonon-assisted defect trapping can be mitigated, which directly results in PL quantum yield (PLQY) increase. (2) The relaxation rate from trapping states to the valance band is expected to be even slower, which may increase the possibility for electron detrapping by ion-electron coupling. (3) The phonon energy coupling into ionic energy reservoir can be slowed down due to suppressed phonon coupling. (4) The activation of ion out of the lattice as mobile ion can be suppressed because each ion in the lattice has less thermal potential, which decreases the PL quenching source. (5) With continuous constant illumination, the PL intensity mostly likely exhibit enhancement, but the details depend on the balance of enhancement effect from localized ion and quenching effect from mobile ion.

Therefore, with decreasing temperature and thus mitigated phonon coupling, many effects in perovskites will be simultaneously impacted, from hot carrier cooling, defect trapping, carrier diffusion and recombination, optical-acoustic phonon relaxation, acoustic phonon-lattice relaxation, and also the ion-electron coupling. Now it is impossible to quantitatively analyze the detailed dynamic processes and therefore, it is very difficult to ascribed to or exclude the observed effect to the proposed ion-electron coupling or any other individual effect, either illumination induced PL enhancement or PL quenching.

## 6. Discussion

It has been well accepted that halide perovskites have the feature of mixed ionic-electronic conduction, and many anomalous phenomena are considered closely relevant to ionic effects. The apparent effect of electron detrapping through ion-electron coupling is the significantly prolonged carrier lifetime and the suppressed energy loss. In other words, the observed carrier lifetime does not match the intrinsic carrier lifetime that is determined by interaction strength of electron-hole interaction and quality of the materials, which are widely regarded as the fundamental mechanism. The carrier recombination processes can be significantly changed by ionic influence. One of the most significant ionic effects is the defects as recombination centres thus the carrier detrapping can be further correlated to other defect relevant factors, such as grain boundary, composition and fabrication, operating temperature, illumination and electric field bias.[38, 87, 143-144] The electron detrapping can directly impacts the performance of the devices, such as defect tolerance and light soaking observed in perovskite solar cells.[40, 74]

The rate equations (4-6) establish the correlation between carrier recombination dynamics in nanoseconds and ionic influence in seconds. In the equations charge carriers exhibit recombination dynamics intrinsically in nanosecond timescale that is determined by the defect trapping and bimolecular recombinations. The ionic contribution through ionic-electronic coupling exhibits a variation in macroscopic timescale, which is closely relevant to the conditions of illumination or electric field, as well as the accumulation time. Therefore, the energetic ions exert their contribution on the electron recombination dynamics through carrier detrapping and results in a slow variation in seconds timescale. These variations are generally sensitive to the intrinsic conditions of composition and fabrication as well as extrinsic conditions of illumination and electronic field.

It should be noted that the observed PL enhancement may be complicated due to the multiple species and multiple roles of ions in perovskites. Mobile ion induced PL quenching has been clearly observed in perovskite nano- or micro-particles as fluorescence intermittency [62, 145], and in bulk as fluorescence quenching[38-39]. Similar illumination-induced fluorescence quenching has also been observed at the interface between perovskite and the hole transport layer (HTL) in either bulk or microscopic samples, ascribed to mobile ion accumulation.[44, 91] PL quenching induced either by illumination or external electric field and phase segregation caused by illumination induced halide substitution have been reported, which strongly suggest the quencher is negatively charged and migratable, most likely the interstitials of $I^-$ and $Br^-$.[46, 59, 80] PL quenching is only observed near the anode interface of $MAPbI_3$/HTL (spiro-OMeTAD), but not at the cathode interface of $MAPbI_3$/ETL (PCBM), Fig. S1, further supporting this argument.[91] The possibility is therefore excluded that halide interstitials is responsible for enhanced carrier detrapping by ion-electron coupling. Moreover, in electric field induced PL imaging, [38] with continuous illumination and the accumulation of negative ions, the fluorescence near the anode is gradually quenched (Fig. S2).

Considering such slow illumination-induced fluorescence variation can be observed in different compositions, inorganic or organic, carrier detrapping effect is speculated to originate from synergy of halide ions. In the same sample, either polycrystalline or single crystal, PL enhancement can be observed at lower excitation intensity but fluorescence quenching at high excitation density[38]. This may suggest the ion-electron coupling forms at lower excitation intensity before halide interstitial is well activated/generated. To date, there is not any species of ions has been confirmed to be directly correlated with PL enhancement, strongly suggest the photobrightening effect is not ascribed to any charged ion, but the ion localized in the lattice. It is still challenging to determine the detailed ion-phonon and electron-ion interactions in halide perovskites, given the high complexity of ions and lack of suitable probe techniques so far.

**Conclusions**

By introducing new concepts of ionic energy reservoir and ion-electron coupling, and in-depth analyzing existing data on the mysterious long lifetime and diffusion length, we propose a hypothesis that efficient electron detrapping through enhanced ion-electron coupling is the fundamental cause of the superior optoelectronic properties of halide perovskites. The intriguing feature of halide perovskites is their mixed ionic-electronic conduction, which is fundamentally different from conventional semiconductors. It is reasonable to suppose ion-electron coupling will inherently impact the optoelectronic properties, which can be reflected in their carrier recombinations. We have elucidated that electron detrapping by ion-electron coupling can directly result in prolonged carrier lifetime and decreased defect trapping, which further correlates with enhanced fluorescence efficiency, increased carrier diffusion length. With such carrier detrapping through ion-electron coupling, carrier dynamics in ns timescale and ion dynamics in timescale of second to hour are well correlated. Differential equations have been established to include the mixed ionic-electronic contribution, in which the missing ionic influence is clearly expressed on the carrier dynamics in timescale of second to hour through external illumination or electric field. The anomalous phenomena of defect healing/curing, defect tolerance and light soaking are consistently interpreted in perovskite materials and devices. Such superior optoelectronic properties of halide perovskites are holistically considered as the essential cause for the high-performance photovoltaics and optoelectronics. This concept provides deep physical understanding in the underlying effect of ionic-electronic interaction, which opens new perspectives in finding the origin for the outstanding optoelectronic properties in halide perovskites. It will also result in new strategies and profound impact to improve the performance and stability of perovskite devices by enhancing ion-electron coupling.

**Acknowledgement**

Authors acknowledge the contribution of Weijian Chen, Mehri Ghasemi, Yang Liu, Chunhua Zhou, Sheng Chen, Rui Sheng, Xiaofan Deng, Qi Li, Junlin Lu, through experimental samples and data acquisition of spectroscopic experiments. **Funding**: Authors acknowledges the financial support from Australian Research Council through Future Fellowship (FT210100806), the Discovery Project (DP220100603), the Centre of Excellence Program (CE230100006) and the Industrial Transformation